\documentstyle[epsfig]{aipproc}

\begin{document}
\thispagestyle{empty}

\phantom{X} \vspace{2.5cm}

\begin{center}
{\Large\bf The polarized Drell--Yan process}
\vskip .1cm
{\Large\bf at ${\cal O}(\alpha_s)$}
\vskip 1.cm
{\large  T.~Gehrmann} 
\vskip .4cm
{\it DESY, Theory Group, D-22603 Hamburg, Germany}
\end{center}
\vskip 3.cm

\begin{center}
{\bf Abstract}
\end{center}
The results of a recent calculation of the QCD corrections to the
polarized Drell--Yan process are summarized. Some implications for the 
production of vector bosons at RHIC are discussed.

\vspace{6.5cm}
\noindent
{\it Talk presented at the  5th International Workshop on 
``Deep Inelastic Scattering and QCD'' 
(DIS '97), Chicago, Illinois, USA, April 14-18, 1997.}
\vfill

\setcounter{page}{0} 
\newpage

\title{The polarized Drell--Yan process\\at ${\cal O}(\alpha_s)$}

\author{T.~Gehrmann}
\address{DESY, Theory Group, D-22603 Hamburg, Germany}

\maketitle

\begin{abstract}
The results of a recent calculation of the QCD corrections to the
polarized Drell--Yan process are summarized. Some implications for the 
production of vector bosons at RHIC are discussed.
\end{abstract}

The production of lepton pairs in hadron collisions, the Drell--Yan 
process~\cite{drellyan}, is one of the most powerful tools to probe the 
structure of hadrons. Its parton model interpretation is straightforward 
--~the process is induced by the annihilation of a quark--antiquark
pair into a virtual photon which subsequently decays into a lepton pair. 
The Drell--Yan process in proton--proton or proton--nucleus collisions
therefore provides a direct probe of the antiquark densities in protons 
and nuclei. Experimental data on the Drell--Yan process in unpolarized 
collisions are crucial to constrain the behaviour of the sea quark 
distributions at large $x$ and to determine the flavour structure of 
the light quark sea. It is therefore natural to expect that a measurement
of the Drell--Yan cross section in polarized hadron--hadron collisions
will yield vital information on the polarization of the quark sea 
in the nucleon, which is presently only poorly constrained~\cite{fits}
from deep inelastic scattering data. 

Apart from the invariant mass distribution, one ususally studies the 
distribution of the lepton pairs as function of the Feynman parameter 
$x_F$ or of the hadron--hadron centre-of-mass rapidity $y$. The resulting 
distributions at fixed invariant mass can be directly related to the  
$x$-dependence of the parton distributions in beam and target. Moreover,
most fixed target experiments have only a limited kinematic coverage in
$x_F$ or $y$, such that only these distributions can be measured without 
extrapolation into experimentally inaccessible regions.

The rather large QCD corrections to the unpolarized Drell--Yan cross 
section~\cite{unpol} suggest that a reliable interpretation of the 
Drell--Yan process in terms of partonic distribution function is only 
possible if higher order corrections are taken into account. Following
closely the method of the unpolarized calculation~\cite{unpol}, we have 
recently calculated~\cite{dypol} 
the next-to-leading order corrections to the $x_F$- and $y$-distributions 
of lepton pairs produced in collisions of longitudinally polarized hadrons. 

Truncated up to ${\cal O}(\alpha_s)$, 
the cross section for the polarized Drell--Yan process 
receives contributions from the 
$q\bar q$--annihilation process at leading and next-to-leading order 
and the quark-gluon Compton scattering process. It can be 
expressed as:
\begin{eqnarray}
\frac{{\rm d}\Delta \sigma}{{\rm d}M^2 {\rm d}x_F} & = & \frac{4\pi \alpha^2}{9 M^2 S} 
\sum_i e_i^2 \int_{x_1^0}^1 {\rm d}x_1 \int_{x_2^0}^1 {\rm d}x_2 \nonumber \\
& & \hspace{-0.3cm}
\times \Bigg\{\left[\frac{{\rm d}\Delta \hat{\sigma}_{q\bar{q}}^{(0)}}
{{\rm d}M^2 {\rm d}x_F} (x_1,x_2) +\frac{\alpha_s}{2\pi}
\frac{{\rm d}\Delta \hat{\sigma}_{q\bar{q}}^{(1)}}
{{\rm d}M^2 {\rm d}x_F}\left(x_1,x_2,\frac{M^2}{\mu_F^2}\right)\right]
\nonumber \\
& & \hspace{0.5cm}
\Big\{ \Delta q_i(x_1,\mu_F^2)\Delta\bar{q}_i(x_2,\mu_F^2) 
+\Delta\bar{q}_i(x_1,\mu_F^2)\Delta q_i(x_2,\mu_F^2) \Big\} \nonumber \\
& & \hspace{0.1cm} + \Bigg[ \frac{\alpha_s}{2\pi}
\frac{{\rm d}\Delta \hat{\sigma}_{qg}^{(1)}} {{\rm d}M^2 {\rm d}x_F} \left(x_1,x_2,
\frac{M^2}{\mu_F^2}\right)\nonumber \\ 
& & \hspace{0.5cm}
\Delta G(x_1,\mu_F^2) 
\left\{ 
\Delta q_i(x_2,\mu_F^2) + 
\Delta \bar{q}_i (x_2,\mu_F^2) \right\} 
+ (1 \leftrightarrow 2)\Bigg] \Bigg\},
\label{eq:xfmaster}
\end{eqnarray}
where analytic expressions for the parton level cross sections in the 
$\overline{{\rm MS}}$--scheme 
are listed in~\cite{dypol}. The 
expression for the $y$-distribution of the Drell--Yan pairs takes a 
similar form. Integration over $x_F$ yields the Drell--Yan 
mass distribution which agrees with earlier results~\cite{check}.
\begin{figure}[b!] 
\centerline{\epsfig{file=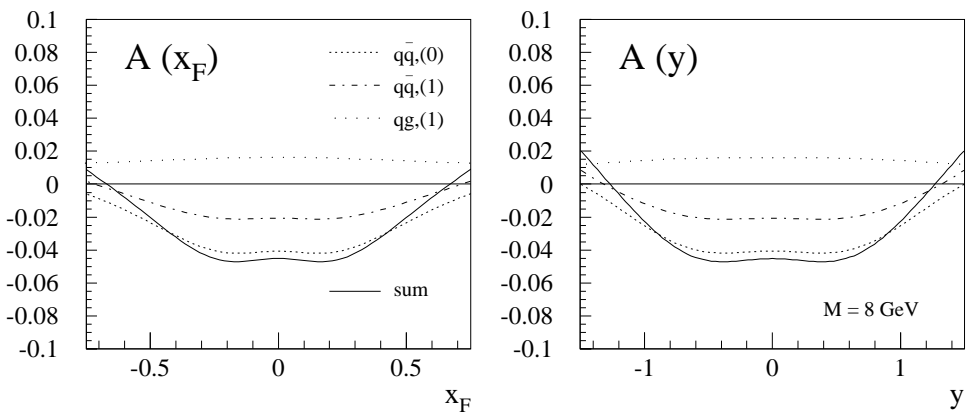,angle=-90,width=13cm}}
\vspace{10pt}
\caption{Contributions of the individual parton level subprocesses 
to the polarized Drell--Yan cross section (see text). }
\label{fig1}
\end{figure}

The numerical importance of these different contributions 
is illustrated in Figure~1, which shows the ratio between polarized and 
unpolarized Drell--Yan cross sections for the collision of a proton 
beam ($E_p=820$~GeV) on a fixed proton target~\cite{heran}. 
All curves are obtained with the polarized GS(A) parton densities~\cite{gs}
and are shown for $M=8$~GeV. The polarized 
subprocess contributions are normalized to the {\it full} unpolarized 
cross section at next-to-leading order. The relative magnitude of the 
individual contributions is similar to the unpolarized Drell--Yan 
process~\cite{unpol}.
The ${\cal O}(\alpha_s)$ correction to the $q\bar q$--annihilation process 
enhances significantly the lowest order prediction while the quark--gluon 
Compton process contributes with a sign opposite to the annihilation process. 
However, the relative magnitude of annihilation and Compton process depends 
on the magnitude of the gluon distribution at large $x$, which is 
completely undetermined at present~\cite{gs,grsv}. This uncertainty 
prevents the sensible prediction of a
$K$--factor between the cross sections at leading and
next-to-leading order.
\begin{figure}[b!] 
\centerline{\epsfig{file=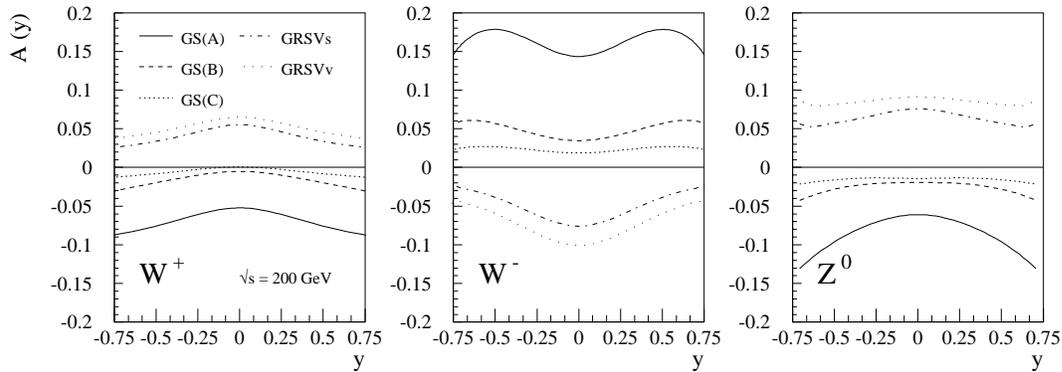,angle=-90,width=14cm}}
\vspace{10pt}
\caption{Expected asymmetries in the production of vector bosons at 
RHIC ($\sqrt s=$200~GeV).}
\label{fig2}
\end{figure}
\begin{figure}[b!] 
\centerline{\epsfig{file=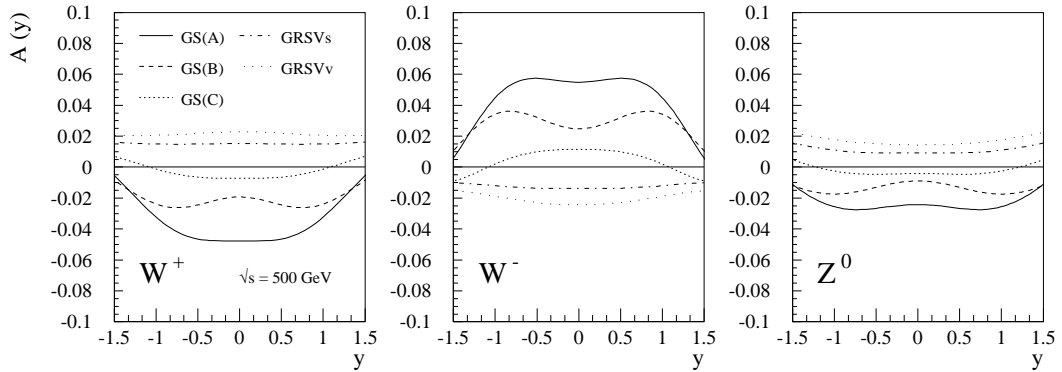,angle=-90,width=14cm}}
\vspace{10pt}
\caption{Expected asymmetries in the production of vector bosons at 
RHIC ($\sqrt s=$500~GeV).}
\label{fig3}
\end{figure}

All above results on the Drell--Yan process 
can immediately be applied to the production of massive vector bosons at 
the polarized proton--proton collider RHIC~\cite{rhic}, which has 
so far only been studied at leading order~\cite{old}. Measuring the 
production asymmetries for $W^{\pm}$ and $Z^{0}$ bosons at RHIC will 
in principle allow to determine the sea quark polarization at large $x$,
a discrimination between $W^+$ and $W^-$ will moreover allow for a 
flavour decomposition of the polarized quark sea, as $W^+$ mainly originate 
from $u\bar d$--annihilation, whereas $W^-$ are produced via 
$d\bar u$--annihilation.

Using different up-to-date parametrizations~\cite{gs,grsv}
of polarized parton distributions,
it is possible to estimate the size of the asymmetry which should be 
expected in the production of vector bosons at RHIC. Figures~2 
and~3 show these asymmetries (obtained in the small width 
approximation, e.g.~\cite{eswbook})
for the different centre-of-mass energies 
($\sqrt s=200$~GeV and $\sqrt s=500$~GeV) planned for RHIC.  It can be
seen in these figures that the production asymmetries for vector 
bosons at RHIC are sizable (around 10\% at 200~GeV and around 5\% at 
500~GeV) and that the different parametrizations yield significantly 
different predictions -- reflecting the present lack of knowledge on the 
polarized sea quark distributions at large $x$. The larger asymmetry at 
200~GeV may suggest a measurement at this energy to be favourable; it should 
however be kept in mind that the vector boson production cross sections 
at this energy are several hundred times smaller than at 500~GeV. 

In summary, the complete ${\cal O}(\alpha_s)$ 
corrections to the $x_F$- and $y$-dependence of the longitudinally polarized  
Drell--Yan cross section 
have been derived recently~\cite{dypol}. These corrections are quantitatively 
similar to the corrections in the unpolarized case and hence sizable even
at collider energies. They
enable a consistent next-to-leading order 
determination of the polarization of sea quarks in the 
nucleon from future measurements of lepton pair production at fixed 
target energies or from massive vector boson production at RHIC.

\end{document}